\documentclass[%
preprint,
10pt,
twocolumn,
nobibnotes,
 amsmath,amssymb,
 aps,
prl]{revtex4-1}
\usepackage{graphicx}
\usepackage{bm}

\newcommand{\dif}{\mathrm{d}}
\newcommand{\xib}{\bar{\xi}}
\newcommand{\etab}{\bar{\eta}}
\newcommand{\epsr}{\epsilon}

\newcommand{\unitr}{\hat{\bm{r}}}
\newcommand{\unitt}{\hat{\bm{\theta}}}
\newcommand{\unitf}{\hat{\bm{\phi}}}
\newcommand{\unitz}{\hat{\bm{z}}}
\newcommand{\unitx}{\hat{\bm{x}}}

\begin{document}


\title{Spheroidal harmonic expansions for the solution of Laplace's equation for a point source near a sphere}

\author{Matt R. A. Maji\'c} 
\author{Baptiste Augui\'e}
\author{Eric C. Le Ru} \email{eric.leru@vuw.ac.nz}

\affiliation{The MacDiarmid Institute for Advanced Materials and Nanotechnology,
School of Chemical and Physical Sciences, Victoria University of Wellington,
PO Box 600, Wellington 6140, New Zealand}

\date{\today}

%

\begin{abstract}
We propose a powerful approach to solve Laplace's equation for point sources near a spherical object. The central new idea is to use prolate spheroidal solid harmonics, which are separable solutions of Laplace's equation in spheroidal coordinates, instead of the more natural spherical solid harmonics. We motivate this choice and show that the resulting series expansions converge much faster. This improvement is discussed in terms of the singularity of the solution and its analytic continuation. The benefits of this approach are illustrated for a specific example: the calculation of modified decay rates of light emitters close to nanostructures in the long-wavelength approximation. We expect the general approach to be applicable with similar benefits to a variety of other contexts, from other geometries to other equations of mathematical physics.
\end{abstract}


\maketitle

Laplace's equation is one of the most important partial differential equations of physics and engineering. It arises in many fields including electromagnetism, classical gravity, and fluid dynamics. It also has close links, through the Laplacian operator, with other important differential equations of physics, such as the wave equation and the diffusion equation. Analytical solutions of Laplace's equation, typically obtained via the method of separation of variables, are standard materials for physics textbooks \cite{1953Morse}.
The solution for a point source located outside a sphere plays a specially important role through its connection
with the Green's function formalism \cite{1941Stratton}.
We will focus on electrostatics in this article, but our results naturally extend to other applications of Laplace's equation.

The standard electrostatics solution for a point source outside a dielectric sphere is relatively straightforward and obtained
as a multipole expansion (infinite series) \cite{1941Stratton,1984FordPR}.
One important and often overlooked property of those series is that they can be very slowly
convergent for sources close to the surface (often the most relevant situation), as shown explicitly in Ref. \cite{2009Book}.
Moroz recently revisited this problem by focusing specifically on the decay rates (i.e. the self-field of a dipole
in the quasi-static approximation)
and used mathematical manipulations to express those series in a more convergent form \cite{2011MorozJPCC}.
Lindell also approached this problem from the point of view of image theory \cite{1992Lindell}, but the resulting
solutions involve integrals which must be computed numerically.

In this work, we propose and demonstrate an alternative approach based on the use of spheroidal harmonics, which are the separable solutions of Laplace's equation in spheroidal coordinates \cite{1941Stratton,1953Morse}.
This choice may appear counter-intuitive for a spherical object,
but the point source breaks the spherical symmetry and we will show that the spheroidal harmonics are better suited to
account for the singularities of the solution. With this original approach, we demonstrate dramatic improvements for the convergence
of the solution series.
We show that this idea is directly applicable to different types of point sources and argue that its applicability could extend to other geometries and other equations of mathematical  physics.

\begin{figure}[h]
\includegraphics[width=7cm,clip=true,trim=6.3cm 7.0cm 4.5cm 0.5cm]{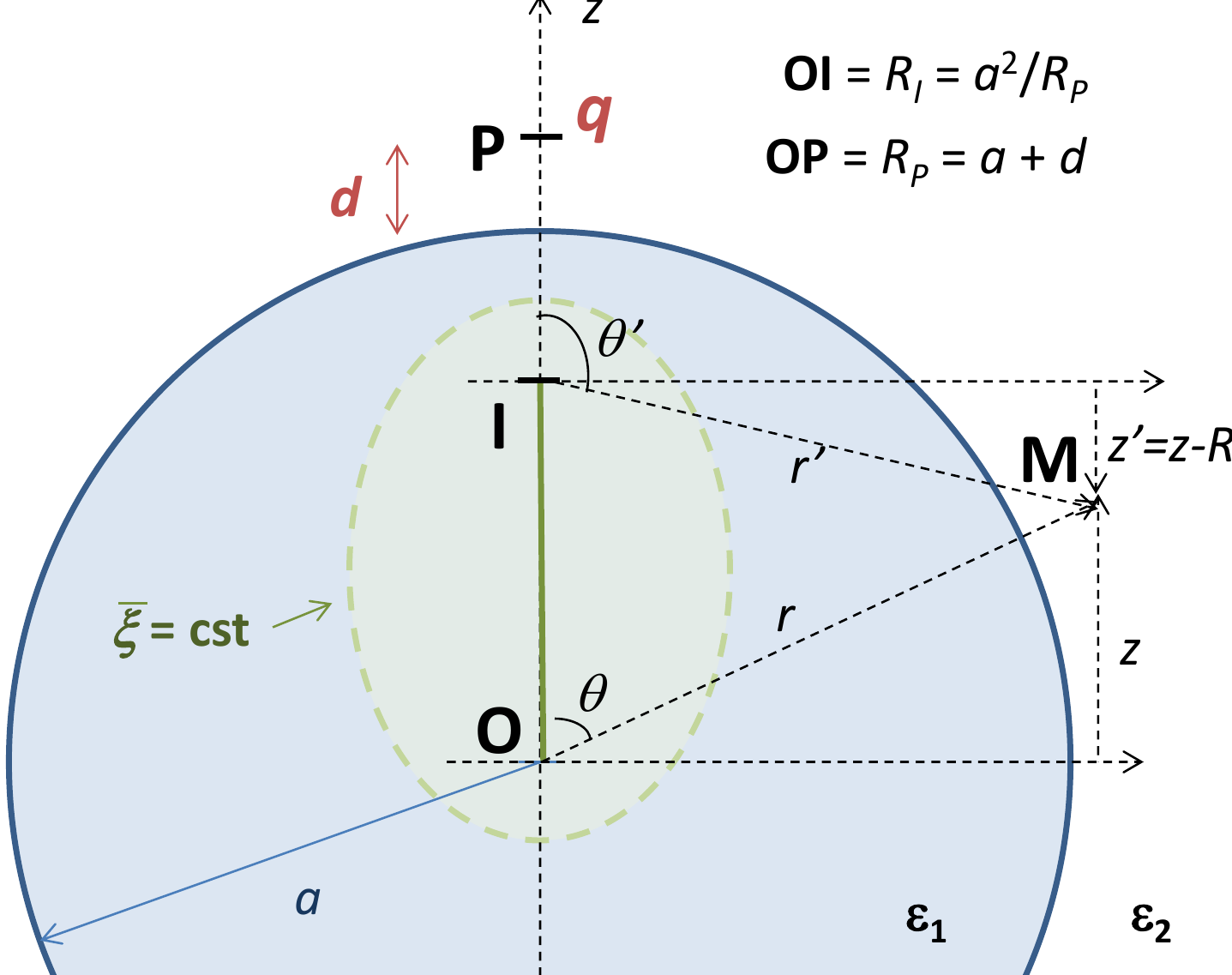}
\caption{Schematic of the electrostatics problem under study: a point charge $q$ at a distance $d$ from a sphere of radius $a$.
The various coordinate systems used in the solution are also illustrated: spherical ($r$,$\theta$,$\phi$), offset
spherical ($r'$,$\theta'$,$\phi$), and offset prolate spheroidal ($\xib$,$\etab$,$\phi$).} 
\label{FigSchem}
\end{figure}

\begin{figure*}
\includegraphics[width=12cm,clip=true,trim=0.5cm 0cm 2.5cm 0cm]{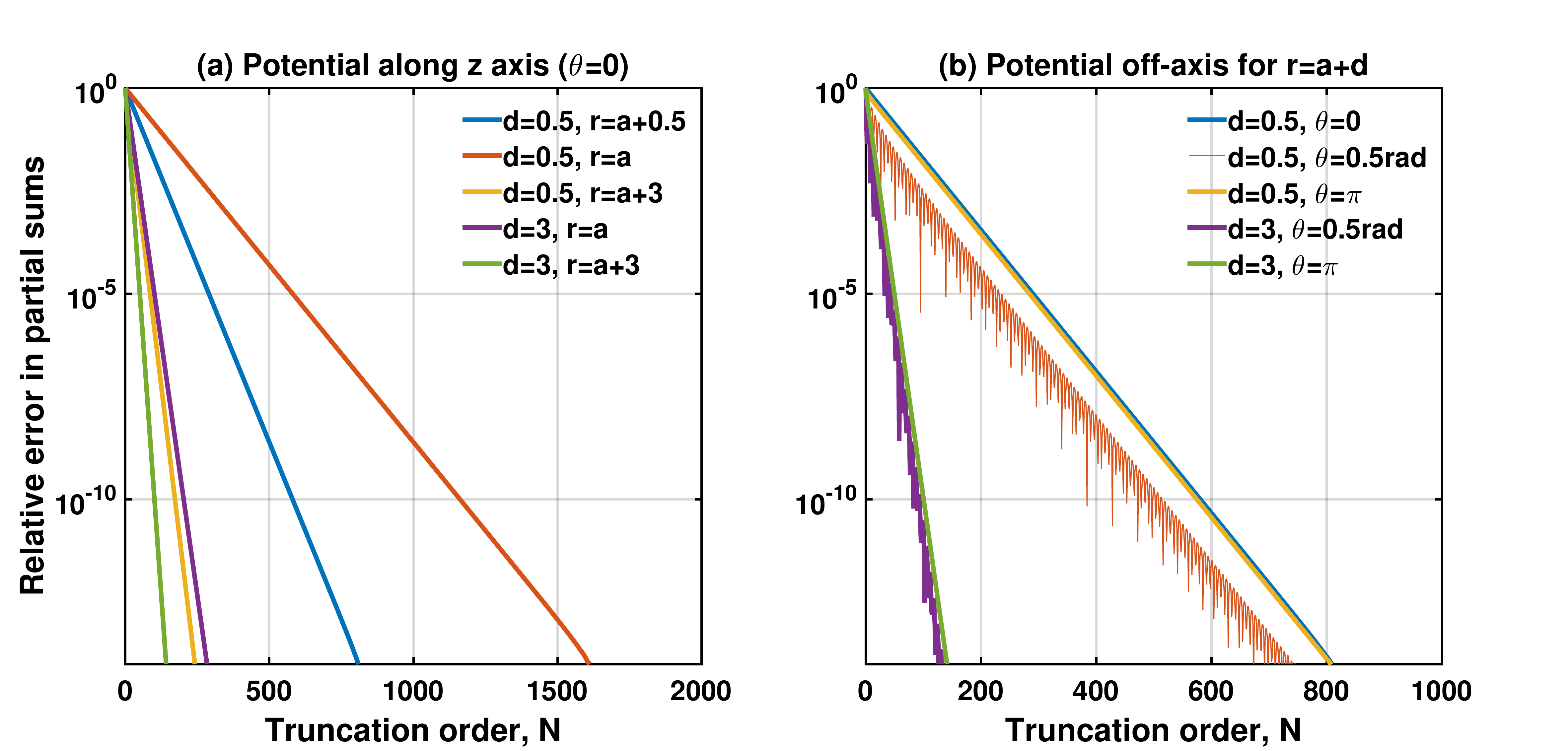}
\caption{Convergence of the standard series solution (Eq. \ref{EqnPhiSca1}) for a point charge 
at a distance of $d=0.5$\,nm or $d=3$\,nm
from an $a=25$\,nm-radius sphere (with $\epsr=-6.5+0.67i$).
The relative errors in the partial sums (with respect to the converged sums) are shown for the outside potential
at different positions either along the $z$-axis (a) or off-axis for $r=a+d$ (b).}
\label{Fig2}
\end{figure*}

To present our new approach, we will first focus on the simplest case of a point charge.
As illustrated in Fig.~\ref{FigSchem}, we consider a point charge $q$ located at $\mathbf{R}_P$, on the $z$-axis at a distance $d$ from a sphere of radius $a$ ($|\mathbf{R}_P| = R_P = a+d)$.
The dielectric relative permittivities of the sphere and embedding medium are $\epsilon_2$ and $\epsilon_1$ respectively
and their ratio is denoted $\epsr={\epsilon_2}/{\epsilon_1}$ for convenience.
Our results will be illustrated for $\epsr={-6.5 + 0.67i}$ (corresponding to a gold sphere in water excited with $\lambda=633$\,nm light in the quasi-static approximation), but similar conclusions were obtained with other values of $\epsr$, including for absorbing or non-absorbing dielectric spheres.
We also choose for illustration a sphere radius of $a=25\,$nm and a distance from the surface of $d=0.5$ or $d=3$\,nm, but note that the results are scale-invariant.

We seek the outside potential $\phi(\mathbf{r})$, solution of Laplace's equation in the presence of this source term.
For convenience, we write $\phi = \bar{\phi} q/(4\pi\epsilon_0\epsilon_1 a) $ and work with the dimensionless $\bar{\phi}$.
The standard solution of this problem consists in expanding the point charge potential $\bar{\phi}_q$ as a series of regular solid harmonics
centered on the sphere \cite{1941Stratton}: 
\begin{align}
\bar{\phi}_q =  \frac{a}{R_P}\sum_{n=0}^\infty \frac{r^n}{R_P^n} P_n(\cos\theta) \quad (r<R_P)
\end{align}
where $(r,\theta,\phi)$ are spherical coordinates and $P_n$ are the Legendre polynomials.
The potential outside the sphere ($r>a$) is then given by $\bar{\phi}_\mathrm{out} = \bar{\phi}_q + \bar{\phi}_r$, with the ``reflected'' potential \cite{1941Stratton}:
\begin{align}
\bar{\phi}_r = -\sum_{n=0}^\infty ~  \beta_n \left(\frac{R_I}{r}\right)^{n+1}  P_n(\cos\theta), 
\label{EqnPhiSca1}
\end{align}
where $R_I=a^2/R_P$ and the adimensional sphere polarizabilities are:
\begin{align}
\beta_n = \frac{n(\epsr-1)}{n(\epsr+1)+1}.
\end{align}
We also define $\beta_\infty = (\epsr-1)/(\epsr+1)$, which is related to the response of a planar interface.

\begin{figure*}
\includegraphics[width=13.5cm,clip=true,trim=0.3cm 6.2cm 0.8cm 1cm]{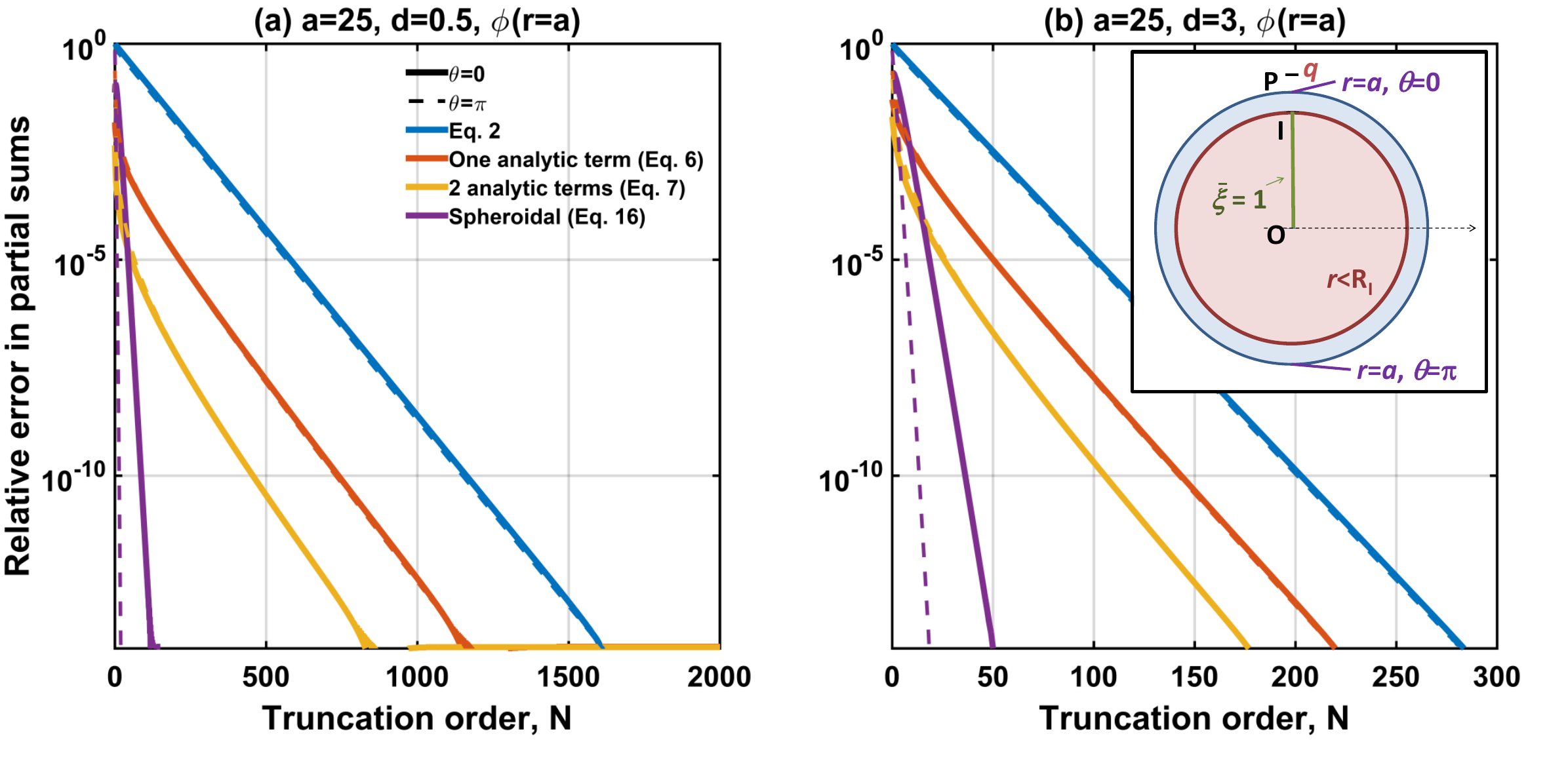}
\caption{Convergence of the improved series solutions for the outside potential at the surface ($r=a$) either close to the source ($\theta=0$, solid lines) or at the opposite side of the sphere ($\theta=\pi$, dashed lines) for a point charge at a distance $d=0.5$\,nm (a) or $d=3$\,nm (b).
We compare the standard solution (Eq. \ref{EqnPhiSca1}) with
the improved solutions with the image charge term (Eq. \ref{EqnPhiSca2}) and with the logarithmic term (Eq. \ref{EqnPhiSca3}). The new approach using a spheroidal harmonics expansion (Eq. \ref{EqnPhiSca6}) is also compared to those and converges much faster, especially for sources very close to the sphere. The inset in (b) depicts the region of divergence of the series for spherical (red) and spheroidal (green) harmonics expansions.}
\label{Fig3}
\end{figure*}

As discussed in Refs. \cite{1992Lindell,2009Book,2011MorozJPCC}, the sum in Eq. \ref{EqnPhiSca1} can be very slowly convergent when evaluated at or in the vicinity
of the sphere surface ($r \approx a$) for a point source close to the sphere ($d\ll a$). This is shown explicitly in Fig. \ref{Fig2} where we computed the relative errors from the partial series of the potential at different points close to or on the
sphere surface. One for example needs to sum more than 1500 terms in the series
to obtain a converged solution (within the double-precision accuracy of $\sim 10^{-15}$) of the potential on the sphere surface when $a/d = 50$. This slow convergence also occurs everywhere on the sphere surface, not just in the vicinity of the point source.
In order to motivate our choice of working with prolate spheroidal coordinates, we first derive a more convergent formulation of the solution with spherical coordinates, where the nature of the singularities of the solution becomes more apparent.
For this, we start from Eq. \ref{EqnPhiSca1}, and isolate the dominant contribution for large $n$ by writing:
\begin{align}
\beta_n &= \beta_\infty -\frac{\beta_\infty}{n(\epsr+1)+1}.
\end{align}
Substituting back into Eq. \ref{EqnPhiSca1}, the second term gives a series that converges faster and the the first term gives a (still slowly-converging) series for which we recognize a closed-form analytical expression \cite{1941Stratton}:
\begin{align}
-\beta_\infty \sum_{n=0}^\infty ~  \left(\frac{R_I}{r}\right)^{n+1}  ~ P_n(\cos\theta) = -\frac{\beta_\infty R_I}{|\mathbf{r} - R_I\unitz|}.
\end{align}
This can be viewed as the potential created by an image point charge $q_I=-q \beta_\infty (R_I/a)$, located at a distance $R_I$ from the origin on the $z$ axis (point I, see Fig.~\ref{FigSchem}). This is the same image charge location as that used in the method of images to solve the same problem for a perfect conductor
\cite{1941Stratton,1998Jackson}.
The solution then takes the form (the primed coordinates refer to those centered at I):
\begin{align}
&\bar{\phi}_r =
 -\beta_\infty\frac{R_I}{r'} 
+\sum_{n=0}^\infty \frac{\beta_\infty}{n(\epsr+1)+1}\left(\frac{R_I}{r}\right)^{n+1}   P_n(\cos\theta). 
\label{EqnPhiSca2}
\end{align}
The slow convergence of the series in Eq.~\ref{EqnPhiSca1} has been partially removed by isolating and recognizing the analytical expression for the image charge. Nevertheless,  the convergence of the series in Eq.~\ref{EqnPhiSca2} remains slow (Fig. \ref{Fig3}).
This approach can be repeated to further improve the convergence.
Isolating the next term and recognizing its closed-form expression, we obtain after manipulation (see Sec.~S.I.):
\begin{align}
&\bar{\phi}_r =
 - \beta_\infty \frac{R_I}{r'} +  \frac{\beta_\infty}{\epsr+1}\ln\frac{r'-z'}{r-z} \nonumber\\
& + \frac{\epsr \beta_\infty}{\epsr+1}\sum_{n=0}^\infty ~ \left(\frac{R_I}{r}\right)^{n+1}\frac{P_n(\cos\theta)}{(n+1)[n(\epsr+1)+1]} . 
\label{EqnPhiSca3}
\end{align}
As shown in Fig. \ref{Fig3}, the convergence of Eq. \ref{EqnPhiSca3} is again improved, 
but still requires a large number of terms ($\sim 800$) to reach double-precision accuracy.

It is also interesting to note that the second term in Eq.~\ref{EqnPhiSca3} exhibits a logarithmic singularity on the line segment OI; this term can therefore be viewed as an extended image source over this segment. Such a line image charge over OI was also found from a direct analysis of the problem within the method of images \cite{1992Lindell}.
This extended line singularity provides the motivation
for our proposed new approach to the problem.
Instead of using a spherical harmonics expansion, we search instead for a solution in a basis of spheroidal harmonics, namely:
\begin{align}	
\bar{\phi}_r = \sum_{n=0}^\infty a_n Q_n(\xib) P_n(\etab).
\label{EqnPhiSca4}
\end{align}
$Q_n(\xib) P_n(\etab)$ are irregular solid prolate spheroidal harmonics, i.e. they are the standard separable solutions of Laplace's equation
(where there is no $\phi$-dependence) in prolate spheroidal coordinates, with $Q_n(\xib)$ the Legendre functions of the second kind.
$\xib$ and $\etab$ are prolate spheroidal coordinates {\it with focal points
at O, center of sphere, and I, position of the image charge}. The segment OI then corresponds
exactly to $\bar{\xi}=1$. Explicitly, $\xib$ and $\etab$ are:
\begin{align}
\xib=\frac{r+r'}{R_I}, \quad \etab =\frac{r-r'}{R_I}.
\end{align}
The ``bar'' notation is used here to emphasize the fact that prolate spheroidal coordinates
are traditionally defined differently with O at the mid-point between the two foci \cite{1953Morse}.
We choose these coordinates because $Q_n(\xib)$ is then singular exactly on the segment OI (i.e. $\xib=1$), where the singularity of the solution
is expected.

To determine the expansion coefficients $a_n$, we first need to find the expansion for
the irregular spherical solid harmonics $P_n(\cos\theta)/r^{n+1}$ in terms of the irregular prolate spheroidal solid harmonics.
Such expansions can be found in the literature \cite{2000Jansen,2002Russian} in the case where the spherical harmonics center is
in the middle of the focal points used for the spheroidal coordinates. In our case however, the sphere
center corresponds to one of the focal points, so new expressions had to be derived. The details
are provided in Sec. S.II. and we here state the final result:
\begin{align}
&{\left(\frac{R_I}{r}\right)^{n+1}} P_n(\cos\theta) =
 \nonumber\\
& \sum_{k=n}^\infty (-1)^{n+k} \frac{2(2k+1)(k+n)!}{n!^2(k-n)!} ~ Q_k(\xib) P_k(\etab).
\label{EqnSphSph}
\end{align}

One can then substitute this expansion into the original solution (Eq. \ref{EqnPhiSca1}),
swap the order of the sums and relabel the indices $n \leftrightarrow k$,
to obtain the coefficients $a_n$ as:
\begin{align}
a_n = -2(2n+1) \sum_{k=1}^n (-1)^{n+k} \frac{(n+k)!}{k!^2 (n-k)!} \beta_k.
\end{align}
For the problem at hand, it is in fact beneficial to first
isolate the point singularity (image charge) identified earlier, since it
does not exhibit the line singularity found in the spheroidal solid harmonics.
We therefore look for a solution of the form:
\begin{align}
\bar{\phi}_r = -\beta_\infty\frac{R_I}{r'} + \sum_{n=0}^\infty b_n Q_n(\xib) P_n(\etab). 
\label{EqnPhiSca5}
\end{align}
As for $a_n$, the coefficients $b_n$ are obtained by substituting Eq.~\ref{EqnSphSph}
into the series in Eq.~\ref{EqnPhiSca2} and swapping the
order of the sums. We obtain:
\begin{align}
b_n &= \beta_\infty 2(2n+1) c_n,\\[0.2cm]
\mathrm{with~} 
c_n &= \sum_{k=0}^n \frac{(-1)^{n+k}}{k(\epsr+1) + 1} ~ \frac{(n+k)!}{k!^2 (n-k)!}.
\end{align}
This expression is however not suitable for practical computations as large numerical errors appear
in the sum at relatively low $n$ ($\approx 20$), but one can derive the following equivalent expression (see Sec. S.III.):
\begin{align}
c_n = \prod_{k=0}^{n} \frac{\mu-k}{\mu+k}, \quad \mathrm{where~}\mu = \frac{1}{\epsr+1}.
\end{align}
With this expression, $c_n$ can be computed easily by recurrence. The solution for the potential takes the form
\begin{align}
\bar{\phi}_r =
 -\beta_\infty \frac{R_I}{r'} + \beta_\infty \sum_{n=0}^\infty ~ 2(2n+1) c_n Q_n(\xib) P_n(\etab). 
\label{EqnPhiSca6}
\end{align}
The convergence of this series is compared in Fig. \ref{Fig3} to those previously obtained
and the improvements are dramatic. It should be noted that care should be taken in the computation of the Legendre functions
of the second kind, which were computed using a backward recurrence and the modified Lentz algorithm \cite{Qalgorithm}.
In the example of Fig.~\ref{Fig3}(a) ($a=25$, $d=0.5$), full accuracy is obtained 
for the surface field close to the point source with only $\sim 100$ terms instead of $\sim 1600$ for the standard solution.
The benefits are even more dramatic elsewhere near the surface, with only 18 terms needed on the other side of the sphere.


To understand these improvements, we recognize that Eq.~\ref{EqnPhiSca6} provides an analytic continuation of Eq.~\ref{EqnPhiSca3}.
Those infinite series are strictly equivalent in the region where they both converge, but their ranges of convergence
are different: Eq.~\ref{EqnPhiSca3} only converges for $r>R_I$, while Eq.~\ref{EqnPhiSca6} converges everywhere except on the segment OI. One naturally expects that slow convergence of either series will occur near the boundary of its region of convergence.
For spheroidal expansions, the point $r=R_P,\theta=\pi$ is far from the segment of divergence OI (see inset in Fig.~\ref{Fig3}(b)) and the series
therefore converges rapidly. For spherical expansions, this point is very close to the sphere of divergence and convergence is very slow.
The logarithmic term in Eq.~\ref{EqnPhiSca3} and previous studies using the method of images \cite{1992Lindell}
suggest that the analytic continuation of the solution is singular
only on the segment OI and the divergence region cannot be further reduced.
This suggests that the spheroidal solid harmonics (centered on the segment OI) are the most natural basis for
this problem, which explains the better convergence even at the point source position, which is
close to the singularity at I.

These arguments indicate that our proposed approach would be applicable to many related problems.
The method can for example be adapted to apply to the solution inside the sphere (this would take us
too far from the argument and will be discussed elsewhere).
It can also be applied to other types of point sources near a sphere with minor modifications.
We discuss below the results obtained for a dipole
$\mathbf{p}$ located at a distance $d$ from the sphere on the $z$-axis (details of the derivations are provided in Secs. S.IV and S.V.).
The dimensionless potential $\bar{\phi}$ is now defined as $\phi = \bar{\phi} p~/(4\pi\epsilon_0\epsilon_1 a R_P)$ for convenience.
For a perpendicular dipole (oriented along $z$), the reflected potential can be expressed using solid spheroidal harmonics expansions as:
\begin{align}
\bar{\phi}_{\bot} =
& \beta_\infty \frac{R_I^2z'}{r'^3} +  \frac{\epsr \beta_\infty}{\epsr+1}\frac{R_I}{r'}
  \nonumber\\
& - \frac{\epsr \beta_\infty}{\epsr+1}\sum_{n=0}^\infty ~ 2(2n+1) c_n Q_n(\xib) P_n(\etab). 
\label{EqnPhiScaDipPerp}
\end{align}
The first two terms correspond to an image dipole and image point charge respectively, while the series
exhibit a line singularity over the segment OI as before.

\begin{figure}
\includegraphics[width=\columnwidth,clip=true,trim=2.5cm 9cm 2cm 0.2cm]{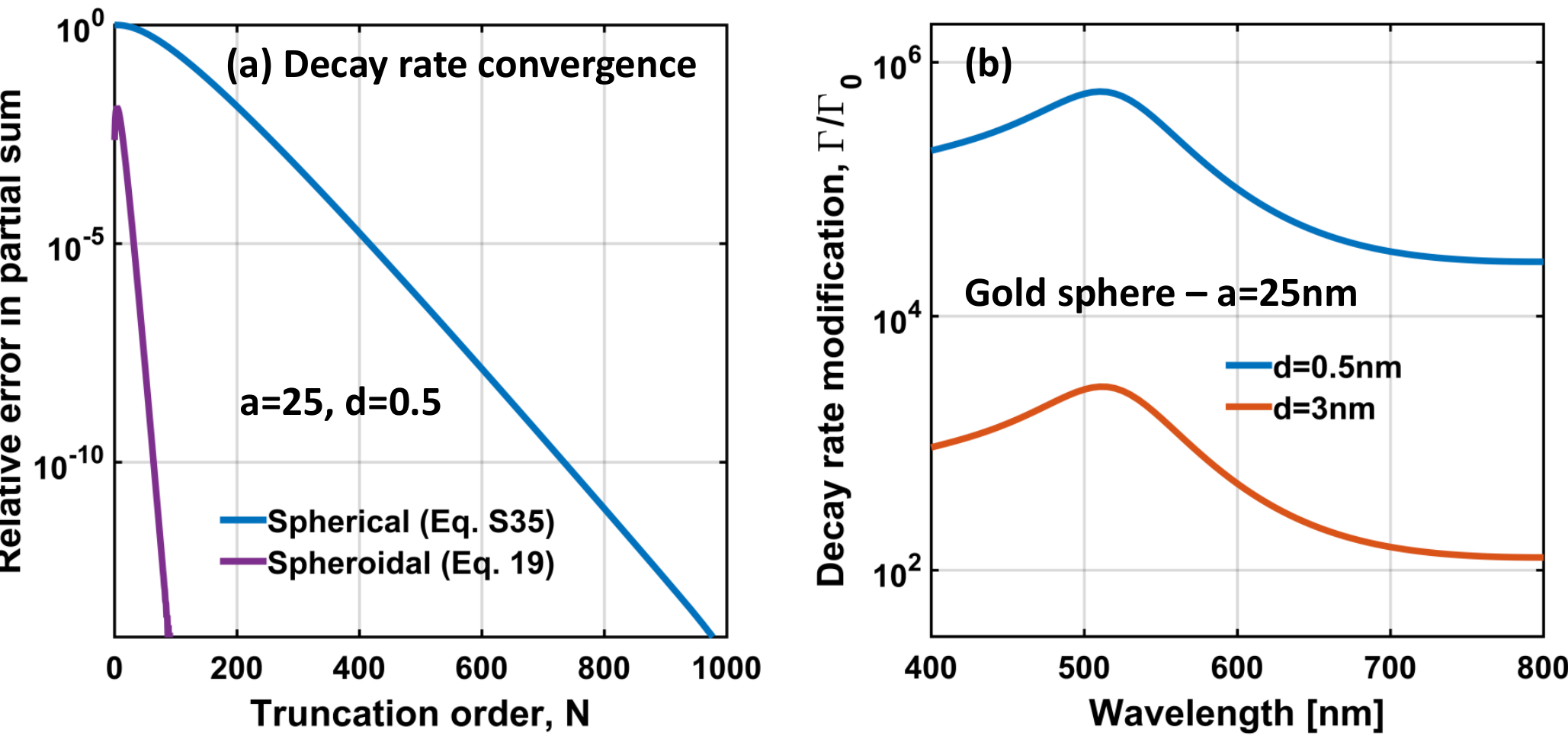}
\caption{(a) Convergence of the standard and improved series solutions for the modified
decay rates $\Gamma_\bot/\Gamma_0$. The spheroidal expansion (Eq.~\ref{EqnMtot2}) requires much fewer terms for accurate results.
(b) Spectral variation of the decay rate modification in the electrostatics approximation for a dipole perpendicular to a gold sphere of radius $a=25$\,nm embedded in water ($\epsilon_1=1.33^2$), at a distance $d=0.5$ or 3\,nm from the surface. The dielectric function for gold is taken from Ref. \cite{2006EtchegoinJCPGold}.}
\label{FigGamma}
\end{figure}

This solution is important in the context of nano-optics, where the modified decay rate $\Gamma$ for a dipolar emitter
can be deduced from its self-field $\mathbf{E}_\mathrm{sf}$ as \cite{1978ChanceACP,1984FordPR,1992BarnettPRL,2006Novotny,2009Book}:
\begin{align}
\frac{\Gamma}{\Gamma_0}=1+\frac{6\pi\epsilon_0\epsilon_1}{k_1^3}
\frac{\mathrm{Im}\left(\mathbf{p}^* \cdot \mathbf{E}_\mathrm{sf}\right)}{|\mathbf{p}|^2},
\end{align}
where $\Gamma_0$ is the normal decay rate in the embedding medium, $k_1=(2\pi/\lambda)\sqrt{\epsilon_1}$ the wave-vector.
In the quasi-static approximation, valid for spheres much smaller than the wavelength, the field solution close to the sphere
can be approximated by the corresponding electrostatics solution. The self-field
$\mathbf{E}_\mathrm{sf}$ can then be obtained by evaluating the reflected field $\mathbf{E}_r = -\boldsymbol{\nabla} \phi_r$
at the dipole position.
For a dipole that is either perpendicular ($\bot$) or parallel ($\parallel$) to the sphere, we obtain (see Secs. S.IV. and S.V.):
\begin{widetext}
\begin{align}
\frac{\Gamma_\bot}{\Gamma_0} &= 1 + \frac{3}{2(k_1 a)^3}\mathrm{Im}\left\lbrace  \beta_\infty \left( \frac{2}{\delta_P^3} + \frac{\epsr}{\epsr+1} \frac{1}{\delta_P^2}
 \right. \right. \nonumber\\ & \left. \left. 
\times \left[ 1 - \frac{1}{\delta_P(1+\delta_P)}\sum_{n=0}^\infty(2n+1)(n+1)c_n[\xib_P Q_n(\xib_P)-Q_{n+1}(\xib_P)] \right]\right)\right\rbrace
\label{EqnMtot2}\\
\frac{\Gamma_\parallel}{\Gamma_0} &= 1 - \frac{3}{2(k_1 a)^3}\mathrm{Im}\left\lbrace \beta_\infty\left(\frac{1}{\delta_P^3} - \frac{2}{\epsr+1}\frac{1}{\sqrt{\delta_P(1+\delta_P)}}\sum_{n=1}^\infty(2n+1)(c_n-1){Q_n^1(\xib_P)}\right)   \right\rbrace.
\end{align}
\end{widetext}
where we have defined $\delta_P = (R_P/a)^2-1$ ($\delta_P \ll 1$ when the dipole is close to the surface), $\xib_P=1+2\delta_P$,
and $Q_n^m$ are the associated Legendre functions of the second kind.
Note that the coefficients $c_n$ in the series depend on $\epsr$ and contribute to the material-dependence of the whole
expression.
In Fig. \ref{FigGamma}(a), the convergence of this new formula is compared for the perpendicular
dipole to that of the standard expression obtained from spherical harmonics expansions \cite{2009Book,2011MorozJPCC}.
Much faster convergence is again evidenced, and almost identical results were obtained for the parallel dipole.
The wavelength dependence of the modified decay rate for a gold nanosphere is shown in Fig. \ref{FigGamma}(b) as an example of application of these new formula.

Beyond those examples, we envisage that our arguments could be extended to other equations of mathematical physics,
such as the Helmholtz equation. The full-wave series solution for a dipole near a sphere are also slowly-convergent but the problem
becomes more acute because numerical problems arise in computing those series for $n$ larger than typically 100 (the terms in the series include
spherical Bessel functions whose values are beyond double-precision arithmetic at large orders). The solutions cannot therefore be easily computed numerically
for dipoles close to the sphere. Spheroidal harmonics expansions (which may involve the standard spheroidal wavefunctions \cite{SpheroidalWF} or alternative definitions) may alleviate such issues.
Spheroidal harmonics expansions could also provide new approaches to solve Laplace's equation with other geometries, for example interacting spheres (where one sphere can be viewed as a source for the other) or of a sphere near an infinite plane. Although those extensions will require further developments, the results presented in this article provide a vivid demonstration of the usefulness of spheroidal coordinates in problems
with symmetry of revolution where they may have been overlooked so far.

%

%
%


\clearpage

\setcounter{equation}{0}
\setcounter{figure}{0}
\setcounter{table}{0}
\makeatletter
\renewcommand{\theequation}{S\arabic{equation}}
\renewcommand{\thefigure}{S\arabic{figure}}
\renewcommand{\bibnumfmt}[1]{[S#1]}
\renewcommand{\citenumfont}[1]{S#1}
\onecolumngrid


{\Large\textbf{Supplimentary Information}}

\author{Matt R. A. Maji\'c} 
\author{Baptiste Augui\'e}
\author{Eric C. Le Ru} \email{eric.leru@vuw.ac.nz}

\affiliation{The MacDiarmid Institute for Advanced Materials and Nanotechnology,
School of Chemical and Physical Sciences, Victoria University of Wellington,
PO Box 600, Wellington 6140, New Zealand}




\tableofcontents

\section{Proof of Eq. 6: closed-form expression for the logarithmic term}

The same approach as carried out to find the image charge term (Eqs. 4-5) can be followed to obtain the next dominant term.
We isolate the next dominant term in $\beta_n$ by writing:
\begin{align}
\frac{n}{n(\epsilon+1)+1}=\frac{1}{\epsilon+1} - \frac{1}{(\epsr+1)^2}\frac{1}{n+1} -\frac{\epsr }{(\epsr+1)^2} \frac{1}{(n+1)[n(\epsr+1)+1]}. \label{splitfracQ}
\end{align}
The choice of factor $1/(n+1)$ in the second term, instead of for example $1/n$, may look arbitrary but will simplify the calculations.
When substituting back Eq.~\ref{splitfracQ} into Eq.~2, the first term gives the same image charge term as found in Eq. 5.
The analytic expression for the sum over the second term in Eq. \ref{splitfracQ} is less obvious.
It involves the series:
\begin{align}
\sum_{n=0}^\infty \left(\frac{R_I}{r}\right)^{n+1} \frac{P_n(\cos\theta)}{n+1}.  
\end{align}
To calculate it, we start from the generating function of the Legendre polynomials:
\begin{align}
\frac{1}{\sqrt{1-2xt+t^2}}=\sum_{n=0}^\infty t^n P_n(x), \qquad (|t|<1).
\end{align}
Integrating with respect to t:
\begin{align}
\ln \left|\frac{t-x + \sqrt{1-2xt+t^2}}{1-x}\right|=\sum_{n=0}^\infty \frac{t^{n+1}}{n+1} P_n(x), \qquad (|t|<1).
\end{align}
Setting $x=\cos\theta$ and $t=R_I/r$, we obtain after simplifications:
\begin{align}
\sum_{n=0}^\infty \left(\frac{R_I}{r}\right)^{n+1} \frac{P_n(\cos\theta)}{n+1}  = \ln \frac{r'-z'}{r-z} \qquad (r>R_I),
\label{EqnLogo0}
\end{align}
where primed coordinates refer to the coordinates in the frame centered on point I:
\begin{align}
\rho'=\rho=r\sin\theta,\quad z' = z-R_I, \quad r'=\sqrt{\rho^2+(z-R_I)^2}=\sqrt{r^2-2zR_I+R_I^2}, \quad \theta'=\cos^{-1}\frac{z'}{r'}.
\end{align}

Eq.~6 then follows for $r>R_I$ by substituting Eq.~\ref{splitfracQ} into Eq.~2 and
using the analytic forms of the series given in Eqs. 5 and \ref{EqnLogo0}.\\

It is also interesting to note that the logarithmic term found here is directly related to the prolate solid spheroidal harmonic for $n=0$:
\begin{align}
Q_0(\bar\xi)P_0(\bar\eta) = \frac{1}{2}\ln\frac{\xib+1}{\xib-1} = \frac{1}{2}\ln \frac{r'-z'}{r-z} \label{Q0}
\end{align}
The latter equality is not obvious but can be proved straightforwardly by inverting the definitions of $\bar\xi$ and $\bar\eta$:
\begin{align}
r=\frac{R_I}{2}(\bar\xi+\bar\eta); \qquad r'=\frac{R_I}{2}(\bar\xi-\bar\eta); \qquad z=\frac{R_I}{2}(\bar\xi\bar\eta+1); \qquad z'=\frac{R_I}{2}(\bar\xi\bar\eta-1)
\end{align}
Then we have
\begin{align}
\frac{r'-z'}{r-z} &= \frac{(\bar\xi-\bar\eta)-(\bar\xi\bar\eta-1)}{(\bar\xi+\bar\eta)-(\bar\xi\bar\eta+1)} 
=\frac{(\bar\xi+1)(\bar\eta-1)}{(\bar\xi-1)(\bar\eta-1)}=\frac{\xib+1}{\xib-1}.
\end{align}
This link (Eq. \ref{Q0}) provides further motivation for the use of spheroidal solid harmonics expansions.

\section{Expansion of spherical solid harmonics in terms of spheroidal solid harmonics}

Such expansions can be found in the literature \cite{2000Jansen,2002Russian} in the case where the spherical harmonics center is in the middle of the focal points used for the spheroidal coordinates. There are four main formulae, corresponding to the expansion of regular (irregular) spherical solid harmonics in terms of regular (irregular) spheroidal harmonics and vice versa. Two of them are relevant to our problem and are reproduced below \cite{2000Jansen,2002Russian}:
\begin{align}
P_n^m(\xi)P_n^m(\eta)&=\sum_{\substack{k=m\\[0.1cm]n+k\text{~even}}}^n~ \frac{(-)^{(n-k)/2}(n+k-1)!!}{(n-k)!!(k+m)!}\frac{(n+m)!}{(n-m)!}\left(\frac{r}{c}\right)^kP_k^m(\cos\theta) 
\\
\left(\frac{c}{r}\right)^{n+1}P_n^m(\cos\theta)&=\sum_{\substack{k=n\\[0.1cm]n+k\text{~even}}}^\infty~\frac{(-)^{(n-k)/2+m}(2k+1)(n+k-1)!!}{(k-n)!!(n-m)!}\frac{(k-m)!}{(k+m)!}
Q_k^m(\xi)P_k^m(\eta) 
\end{align}
where $(-)^n$ is shorthand for $(-1)^n$ and the foci are located at $z=\pm c$ on the $z$-axis and the prolate spheroidal coordinates for those focal points are given by
(we use the definition of Ref. \cite{1953Morse}):
\begin{align*}
\xi=\frac{\sqrt{r^2+2cz+r^2}+\sqrt{r^2-2cz+r^2}}{2c}, \qquad \eta=\frac{\sqrt{r^2+2cz+r^2}-\sqrt{r^2-2cz+r^2}}{2c}.
\end{align*}
In our case however, the sphere center corresponds to one of the focal points and the other one is a $z=R_I$, so the spheroidal coordinates are therefore given as:
\begin{align}
\xib=\frac{r+\sqrt{r^2-2zc+c^2}}{c}, \qquad \etab=\frac{r-\sqrt{r^2-2zc+c^2}}{c}.
\end{align}
with $c\equiv R_I$. We therefore derived new expansions between spherical and the corresponding offset spheroidal harmonics:
\begin{align}
P_n^m(\xib)P_n^m(\etab)=\frac{(n+m)!}{(n-m)!}&\sum_{k=m}^n~\frac{(-)^{n+k}}{k!(k+m)!}\frac{(n+k)!}{(n-k)!}\left(\frac{r}{c}\right)^kP_k^m(\cos\theta) \label{PPnmvsPnm} \\
\frac{c^{n+1}}{r^{n+1}}P_n^m(\cos\theta)=\frac{2}{n!(n-m)!}&\sum_{k=n}^\infty~(-)^{n+m+k}(2k+1)\frac{(k+n)!}{(k-n)!}\frac{(k-m)!}{(k+m)!}Q_k^m(\xib)P_k^m(\etab) \label{PnmvsQPnm}
\end{align}
The relation we use in the manuscript (Eq.~10) is Eq.~\ref{PnmvsQPnm} with $c\equiv R_I$ and $m=0$. It provides an expansion of the irregular solid spherical harmonics in terms of irregular spheroidal harmonics. Eq.~\ref{PPnmvsPnm} is only needed here in the proof of Eq.~\ref{PnmvsQPnm}.
~\\

\noindent\textbf{Proof of Eq. \ref{PPnmvsPnm}}\\

Let us consider the expansion of $P_n^m(\xib)P_n^m(\etab)\exp(im\phi)$ in terms of regular solid harmonics $r^kP_k^m(\cos\theta)\exp(im\phi)$, which must exist since the solid harmonics are a basis for regular solutions to Laplace's equation. We can assume the $m$ are the same on both sides since $\phi$ is the same in both coordinate systems and $\exp(im\phi)$ are linearly independent functions. So we write:
\begin{align}
P_n^m(\xib)P_n^m(\etab) =\sum_{k=0}^\infty~\alpha^m_{nk}\left(\frac{r}{c}\right)^k P_k^m(\cos\theta).
\label{EqnExp1}
\end{align}
The associated Legendre functions (see definitions in Sec. \ref{SecLegDef}) can be written as
\begin{align}
P_n^m(x)=
\begin{cases} 
      (-)^m (1-x^2)^{m/2} \Pi_n^m(x) & |x|<1 \\
      (x^2-1)^{m/2} \Pi_n^m(x) & |x|>1 \\
\end{cases} \qquad \text{, where } \Pi_n^m(x)=\frac{\dif^m }{\dif x^m}P_n(x).
\end{align}
Using the following relation:
\begin{align}
(1-\etab^2)(\xib^2-1) = \frac{4r^2}{c^2}\sin^2\theta
\end{align}
we deduce
\begin{align}
P_n^m(\xib)P_n^m(\etab) = (-)^m \left(\frac{2r}{c}\right)^m \sin^m\theta~ \Pi_n^m(\xib)\Pi_n^m(\etab)
\end{align}
Eq.~\ref{EqnExp1} can therefore be written in terms of $\Pi_n^m$ as
\begin{align}
\Pi_n^m(\xib)\Pi_n^m(\etab) =\sum_{k=m}^\infty~\alpha^m_{nk}\left(\frac{r}{c}\right)^{k-m} \frac{1}{2^m} \Pi_k^m(\cos\theta)\label{EqnExp2}
\end{align}
The coefficients $\alpha^m_{nk}$ can be determined by evaluating the expansion for $\theta=0$ and $r>c,z>c$, which implies $\etab=1$ and $r=c(\xib+1)/2$.
Using the special value:
\begin{align}
\Pi_n^m(1) = \frac{1}{2^m} \frac{(n+m)!}{m! (n-m)!}
\end{align}
we obtain:
\begin{align}
\Pi_n^m(\xib) =\sum_{k=0}^\infty~\alpha^m_{nk} \frac{(k+m)!(n-m)!}{(k-m)!(n+m)!} \frac{1}{2^m} \left(\frac{\xib+1}{2}\right)^{k-m}.
\end{align}
We then use the following identity for the Legendre polynomials
\begin{align}
P_n(x) =\sum_{k=0}^n~(-)^{n+k}\frac{(n+k)!}{k!^2(n-k)!}\left(\frac{x+1}{2}\right)^k.
\end{align}
By differentiating w.r.t $x$, $m$ times, we get
\begin{align}
\Pi_n^m(x) = \sum_{k=m}^n~(-)^{n+k}\frac{(n+k)!}{k!^2(n-k)!}\frac{k!}{2^m(k-m)!}\left(\frac{x+1}{2}\right)^{k-m}.
\end{align}
From this, we identify
\begin{align}
\alpha^m_{nk} = \frac{(n+m)!}{(n-m)!}~(-)^{n+k}\frac{(n+k)!}{k!(n-k)!(k+m)!} \quad \mathrm{for~}m\le k\le n, \quad 0\mathrm{~otherwise,}
\end{align}
which proves Eq. \ref{PPnmvsPnm}.
Note that since the expansion is finite, it converges everywhere and is valid in all space.\\

\textbf{Proof of Eq.~\ref{PnmvsQPnm}}\\

To prove Eq.~\ref{PnmvsQPnm}, we will make use of the expansions of Green's function in terms of both spherical and spheroidal solid harmonics.
The expansion in terms of spheroidal harmonics can be found for example in Ref.~\cite{2000Jansen} for standard prolate spheroidal coordinates and it can be adapted to our modified coordinates with a simple scaling factor of 2, which comes from shrinking the focal length of the coordinates from $2c$ to $c$.
For two points $\mathbf{r}_1$ and $\mathbf{r}_2$ with spheroidal coordinates denoted $(\xib_1,\etab_1,\phi_1)$ and $(\xib_2,\etab_2,\phi_2)$, we have when $\xib_1<\xib_2$:
\begin{align}
\frac{1}{|\mathbf{r}_1-\mathbf{r}_2|}=\sum_{k=0}^\infty 2\frac{2k+1}{c}\sum_{m=0}^k (-)^m(2-\delta_{m0})\frac{(k-m)!^2}{(k+m)!^2}
P_k^m(\xib_1)P_k^m(\etab_1)Q_k^m(\xib_2)P_k^m(\etab_2)\cos m(\phi_1-\phi_2).  
\label{GF1}
\end{align}
We can write a similar expansion with spherical solid harmonics \cite{1998Jackson} when $r_1<r_2$:
\begin{align}
\frac{1}{|\mathbf{r}_1-\mathbf{r}_2|}=\sum_{n=0}^\infty\frac{r_1^n}{r_2^{n+1}}\sum_{m=0}^n (2-\delta_{m0})\frac{(n-m)!}{(n+m)!}P_n^m(\cos\theta_1)P_n^m(\cos\theta_2)\cos m(\phi_1-\phi_2) 
\label{GF2}
\end{align}
where $(r_1,\theta_1,\phi_1)$ and $(r_2,\theta_2,\phi_2)$ are the spherical coordinates of  
$\mathbf{r}_1$ and $\mathbf{r}_2$.
We then substitute Eq.~\ref{PPnmvsPnm} for $\mathbf{r_1}$ into Eq.~\ref{GF1} to express it as an expansion on the
same spherical harmonics basis:
\begin{align}
\frac{1}{|\mathbf{r}_1-\mathbf{r}_2|}&=\sum_{k=0}^\infty 2\frac{2k+1}{c}\sum_{m=0}^k (-)^m(2-\delta_{m0})\frac{(k-m)!^2}{(k+m)!^2}
\left(     \frac{(k+m)!}{(k-m)!}\sum_{n=m}^k~\frac{(-)^{k+n}}{n!(n+m)!}\frac{(k+n)!}{(k-n)!}\frac{r_1^n}{c^n}P_n^m(\cos\theta_1)     \right)\nonumber \\ 
&\hspace{5cm}\times Q_k^m(\xib_2)P_k^m(\etab_2)\cos m(\phi_1-\phi_2).  \nonumber \\[0.5cm]
&\hspace{-2cm}=\sum_{n=0}^\infty \sum_{m=0}^n \sum_{k=n}^\infty  
2\frac{2k+1}{c} (2-\delta_{m0})\frac{(-)^{n+k+m} (k-m)!(k+n)!}{(k+m)!(k-n)!n!(n+m)!}
\frac{r_1^n}{c^n}P_n^m(\cos\theta_1) Q_k^m(\xib_2)P_k^m(\etab_2)\cos m(\phi_1-\phi_2).  
\label{GF3}
\end{align}
where we have swapped the order of the sums using first $\sum_{m=0}^k \sum_{n=m}^k = \sum_{n=0}^k \sum_{m=0}^n$ and then\\
$\sum_{k=0}^\infty \sum_{n=0}^k = \sum_{n=0}^\infty \sum_{k=n}^\infty$.\\
 
Because the functions $r_1^n P_n^m(\cos\theta_1)\cos m(\phi_1-\phi_2)$ are linearly independent, we can equate all terms with same $n$ and $m$ in Eqs. \ref{GF2} and \ref{GF3}
to get:
\begin{align}
\frac{1}{r_2^{n+1}} \frac{(n-m)!}{(n+m)!}P_n^m(\cos\theta_2)=
\sum_{k=n}^\infty 2\frac{2k+1}{c} \frac{(-)^{n+k+m} (k-m)!(k+n)!}{(k+m)!(k-n)!n!(n+m)!}
\frac{1}{c^n}Q_k^m(\xib_2)P_k^m(\etab_2),  
\end{align}
which can be simplified to obtain Eq. \ref{PnmvsQPnm}. This expansion is valid everywhere except on the line segment between the foci at $z=0$ and $z=c$.

%

\section{Simplification of coefficients $c_n$: proof of Eq. 15}

We recall the definition of $c_n$ (Eq. 14) for $n \ge 0$:
\begin{align}
c_n = \sum_{k=0}^n\frac{(-)^{n+k}}{k(\epsr+1)+1}\frac{(n+k)!}{(n-k)!k!^2}  = \mu\sum_{k=0}^n\frac{(-)^{n+k}}{k+\mu}\frac{(n+k)!}{(n-k)!k!^2}
\quad \mathrm{with~} \mu=\frac{1}{\epsr+1}. 
\end{align}
We want to prove that
\begin{align}
\sum_{k=0}^n\frac{(-)^k}{k+\mu}\frac{(n+k)!}{(n-k)!k!^2}
 = \frac{(-)^n}{\mu}\prod_{k=0}^n\frac{\mu-k}{\mu+k} = \frac{(1-\mu)(2-\mu)...(n-\mu)}{\mu(\mu+1)(\mu+2)...(\mu+n)}.
\label{cnproof}
\end{align}
for a general $\mu$.
We first convert the l.h.s into a single fraction:
\begin{align*}
\mathrm{l.h.s}&=\frac{1}{\mu}-\frac{(n+1)n}{\mu+1}+\frac{(n+2)(n+1)n(n-1)}{(\mu+2)2!^2}-\frac{(n+3)(n+2)(n+1)n(n-1)(n-2)}{(\mu+3)3!^2}+... +(-)^n\frac{2n(2n-1)...1}{(\mu+n)n!^2}\\
=&\frac{(\mu+1)...(\mu+n) - (n+1)n\alpha(\mu+2)(\mu+3)...(\mu+n) + \frac{1}{2!^2}(n+2)(n+1)n(n-1)\mu(\mu+1)(\mu+3)...(\mu+n) - ...}{\mu(\mu+1)(\mu+2)...(\mu+n)}
\end{align*}
Both the left and right hand sides of Eq.~\ref{cnproof} are then fractions with the same denominator $\mu(\mu+1)(\mu+2)...(\mu+n)$, and their numerators are polynomials of $\mu$ of degree $n$. These will be equal if they are equal at $n+1$ different points. Define $f(\mu)$ as the l.h.s numerator and $g(\mu)$ as the r.h.s numerator. Choose the $n+1$ points to be at $\mu=-q$ where $q=0,1,2,...n$. For $g(\mu)$ it is easy to show that
\begin{align*}
g(-q)=(1+q)(2+q)...(n+q)=\frac{(n+q)!}{q!}.
\end{align*}
For $f(\mu)$, note that it is composed of a sum of terms of the form $[(-)^b/b!^2]\mu(\mu+1)...(\mu+b-1)(\mu+b+1)...(\mu+n)\times(n+b)(n+b-1)...(n-b+1)$ for some $b\in \{0,1,...n\}$. When you set $\mu=-q$, all terms vanish except the $(q+1)^\mathrm{th}$ one. This term is:
\begin{align*}
f(-q)&=(-)^q\frac{1}{q!^2}(n+q)(n+q-1)...(n-q+1)\times(-q)(-q-1)...(-)\times(1)(2)...(n-q) \\
&=\frac{(n+q)!}{q!}.
\end{align*}
This applies to all $n+1$ values of $q$, so $f(\mu)=g(\mu)$, which proves Eq. \ref{cnproof} (Eq. 15 of the manuscript).

\section{Perpendicular dipole near a sphere}

\subsection{Standard solution in spherical coordinates}

We consider a dipole $\mathbf{p}=p \hat{\mathbf{z}}$ located at $\mathbf{R}_P$, on the $z$-axis at a distance $d$ from a sphere centred at the origin (i.e. $|\mathbf{R}_P| = R = a+d)$.
For convenience, the dimensionless potential $\bar{\phi_\perp}$ is defined as $\phi = \bar{\phi_\perp} p/(4\pi\epsilon_0\epsilon_1 a R_P)$.
The standard solution of this problem consists of expanding the dipole potential $\bar{\phi}_{\mathrm{dip}-{\bot}}$ as a series of regular solid harmonics centered at the origin:
\begin{align}
\bar{\phi}_{\mathrm{dip}-{\bot}}(\mathbf{r}) = aR_P\frac{z-R_P}{|\mathbf{r}-R_P\hat{\mathbf{z}}|^3} = - \frac{a}{R_P}\sum_{n=0}^\infty~ (n+1)\left(\frac{r}{R_P}\right)^n P_n(\cos\theta) \quad (r<R_P).
\label{EqnPhiDip1}
\end{align}
The potential outside the sphere is $\phi_{\mathrm{out}-\bot}=\bar{\phi}_{\mathrm{dip}-{\bot}}+ \bar{\phi}_{\bot}$ with the reflected potential $\bar{\phi}_{\bot}(\mathbf{r})$ \cite{1941Stratton}:
\begin{align}
\bar{\phi}_{\bot} = \sum_{n=1}^\infty ~ (n+1) \beta_n \left(\frac{R_I}{r}\right)^{n+1} P_n(\cos\theta), 
\label{EqnPhiSca1}
\end{align}
where $\beta_n$ are the adimensional sphere polarizabilities defined in Eq. 3.

The electrostatic field can then be obtained from
\begin{align}
\mathbf{E}_{\bot} = -\boldsymbol{\nabla} \phi_{\bot} 
= -\frac{\partial \phi_{\bot}}{\partial r}\hat{\bm{r}} - \frac{1}{r}\frac{\partial \phi_{\bot}}{\partial \theta}\hat{\bm{\theta} } - \frac{1}{r\sin\theta}\frac{\partial \phi_{\bot}}{\partial \phi}\hat{\bm{\phi}} ~~= -E_0 R_I\boldsymbol{\nabla} \bar{\phi}_{\bot} , \qquad \mathrm{with~~}  E_0 = \frac{p}{4\pi\epsilon_0\epsilon_1a^3}.\nonumber
\end{align}
This gives
\begin{align}
\mathbf{E}_{{\bot}}(\mathbf{r}) = E_0 \sum_{n=1}^\infty ~
(n+1) \beta_n \left(\frac{a^2}{R_P r}\right)^{n+2}  \left[
 (n+1) P_n(\cos\theta) \unitr - \frac{\mathrm{d}}{\mathrm{d}\theta}\left[ P_n(\cos\theta)\right]\unitt\right],
\label{EqnESca1}
\end{align}

The self-field at the dipole position ($r=R_P$, $\theta=0$) is then
\begin{align}
\mathbf{E}_{\mathrm{sf}-{\bot}} &= E_0 \sum_{n=1}^\infty ~
(n+1)^2 \beta_n \left(\frac{a^2}{R_P^2}\right)^{n+2}  
 \unitz.
\end{align}

From this, we can use Eq. 18 to deduce the modified decay rate in the electrostatics approximation
\cite{2011MorozJPCC,2009Book}:
\begin{align}
\frac{\Gamma_\bot}{\Gamma_0}=1+\frac{3}{2(k_1 a)^3}
\sum_{n=1} ~ (n+1)^2 \mathrm{Im}(\beta_n) \left(\frac{a^2}{R_P^2}\right)^{n+2}.
\label{EqnMtot1}
\end{align}

\subsection{Analytic expressions for image sources}

We start from the reflected potential for a perpendicular dipole given in Eq.~\ref{EqnPhiSca1}.
Writing $\beta_n$ explicitly, it can be rewritten as: 
\begin{align}
\bar\phi_{\bot}=(\epsr-1)\sum_{n=0}^\infty\frac{n(n+1)}{n(\epsr+1)+1}\left(\frac{R_I}{r}\right)^{n+1}P_n(\cos\theta)
\label{EqnStart}
\end{align}
We then isolate the first two dominant terms in the fraction (as $n\rightarrow \infty$):
\begin{align}
\frac{n(n+1)}{n(\epsr+1)+1}=\frac{n}{\epsr+1}+\frac{\epsr}{(\epsr+1)^2}-\frac{\epsr}{(\epsr+1)^2[n(\epsr+1)+1]} \label{splitfrac2}
\end{align}
The sum over the second term on the r.h.s in Eq. \ref{splitfrac2} is the same as found for the point charge (Eq.~5)
and can be evaluated analytically for $r>R_I$:
\begin{align}
\sum_{n=0}^\infty\left(\frac{R_I}{r}\right)^{n+1}P_n(\cos\theta) = \frac{R_I}{r'}.\label{gf_formula}
\end{align}
This is proportional to the potential created by an image point charge located at I.

Eq.~\ref{gf_formula} may be recognized as the expression for translation of solid spherical harmonics along the $z$ axis \cite{1998Jackson} and can be proved
using the generating function of the Legendre polynomials \cite{Abramovitz}.
A similar expression exists for the sum over the first term. It can be obtained by differentiating Eq. \ref{gf_formula} with respect to $z$.
First note that for $n\ge 0$:
\begin{align}
\frac{\partial}{\partial z}\left[\left(\frac{R_I}{r}\right)^{n+1}P_n(\cos\theta)\right]=-\frac{n+1}{R_I} \left(\frac{R_I}{r}\right)^{n+2}P_{n+1}(\cos\theta)
\label{dbydzSn}
\end{align}
Then by applying $R_I\partial_z$ to both sides of equation \ref{gf_formula} and re-indexing the sum from $n$ to $n-1$ we obtain:
\begin{align}
\sum_{n=0}^\infty n\left(\frac{R_I}{r}\right)^{n+1}P_n(\cos\theta)=\left(\frac{R_I}{r'}\right)^2\cos\theta' = R_I^2\frac{z'}{r'^3} 
\label{EqnOffsetDip}
\end{align}
which is proportional to the potential of an image dipole located at I oriented along $z$ \cite{1998Jackson}, and provides an analytic expression for the sum over the first term on the r.h.s of Eq. \ref{splitfrac2}.
By substituting Eq. \ref{splitfrac2} into Eq. \ref{EqnStart} and using the analytic forms of the series given in Eqs. \ref{gf_formula},\ref{EqnOffsetDip},
we obtain:
\begin{align}
\bar{\phi}_{{\bot}} =
\beta_\infty \frac{R_I^2z'}{r'^3} +  \frac{\epsilon \beta_\infty}{\epsilon+1}\frac{R_I}{r'} - \frac{\epsilon \beta_\infty}{(\epsilon+1)}\sum_{n=0}^\infty ~ \left(\frac{R_I}{r}\right)^{n+1} \frac{1}{n(\epsilon+1)+1} P_n(\cos\theta). 
\label{EqnPhiSca2b}
\end{align}
The same approach can be followed to obtain the next image source by splitting off the next leading order ($1/(n+1)$ dependence) of $\beta_n$. In fact this term is the same logarithmic source that was obtained for the point charge. Because it is singular on the segment OI, we do not separate it and include it in the expansion in terms of spheroidal harmonics.

%

\subsection{New approach with spheroidal harmonics}

Following the same logic as in the manuscript for a point charge, we search for an equivalent solution where we keep
the image point source terms (there are two of them here) and express the rest as a spheroidal solid harmonics expansion:
\begin{align}
\bar{\phi}_{{\bot}} =
& R_I^2\beta_\infty \frac{z'}{r'^3} +  \frac{\epsr \beta_\infty}{\epsr+1}\frac{R_I}{r'}
+\sum_{n=0} ~ d_n Q_n(\xib) P_n(\etab). 
\label{EqnPhiSca5}
\end{align}
The coefficients $d_n$ are again obtained by substituting Eq.~10
into the series in Eq.~\ref{EqnPhiSca2b} and swapping the
order of the sums. We obtain:
\begin{align}
d_n &= -\beta_\infty\frac{\epsr}{\epsr+1} 2(2n+1) \sum_{k=0}^n \frac{(-1)^{n+k}}{k(\epsr+1) + 1} ~ \frac{(n+k)!}{k!^2 (n-k)!}
= -\beta_\infty\frac{\epsr}{\epsr+1} 2(2n+1) c_n,\\
\end{align}
where $c_n$ has been defined in Eqs.~14 and 15.

The new potential solution then takes the form:
\begin{align}
\bar{\phi}_{{\bot}} = \beta_\infty R_I^2\frac{z'}{r'^3} +  \frac{\epsr \beta_\infty}{\epsr+1}\frac{R_I}{r'}
 - 2\frac{\epsr \beta_\infty}{\epsr+1}\sum_{n=0}^\infty ~ (2n+1) c_n Q_n(\xib) P_n(\etab). 
\label{EqnPhiSca6}
\end{align}

\subsection{Electric field and modified decay rate with new formulation}

From this latest expression, we can deduce the corresponding electric field:
For convenience, we provide explicit expressions for the components of the electric field. Note that multiple coordinates can be used ($r$, $\theta$, $r'$, $\theta'$, $\bar\xi$, $\bar\eta$) to express the field differently. We chose for simplicity to keep expressions involving a mixture of those coordinates.
These are:
\begin{align}
E_{\bot}^r &= E_0 \beta_\infty  \left[
\left(2\frac{R_I^3}{r'^3}\cos\theta - 3\frac{R_I^4 r}{r'^5}\sin^2\theta\right)  
-\frac{\epsr}{\epsr+1}\frac{R_I^2}{r'^3}(R_I\cos\theta-r) \right. \nonumber\\
 &\qquad \left.   - \frac{2\epsr}{\epsr+1}\frac{R_I}{r}\sum_{n=0}^\infty (2n+1)(n+1)c_n\left( Q_n(\xib)P_n(\etab) + \frac{Q_n(\xib)P_{n+1}(\etab)-Q_{n+1}(\xib)P_n(\etab)}{\xib-\etab} \right)\right] \nonumber\\[0.5cm]
E_{\bot}^\theta &= E_0 \beta_\infty \sin\theta \left[
 \frac{R_I^3(r^2+R_Iz-2R_I^2)}{r'^5}
+\frac{\epsr}{\epsr+1}\frac{R_I^3}{r'^3}    \right. \nonumber\\
 &\qquad \left.    -\frac{2\epsr}{\epsr+1}\frac{R_I}{r}\sin\theta\sum_{n=0}^\infty (2n+1)(n+1)c_n \left( \frac{Q_n(\xib)P_n(\etab)\cos\theta}{\sin^2\theta} + \frac{r}{r'} \left( \frac{Q_n(\xib)P_{n+1}(\etab)}{\etab^2-1} - \frac{Q_{n+1}(\xib)P_n(\etab)}{\xib^2-1} \right)\right)\right]\nonumber\\[0.5cm]
E_{\bot}^\phi &= 0
\end{align}

The dipole position corresponds to coordinates
\begin{align*}
&r=R_P, &\theta=0,\nonumber\\
&r'= R_P-R_I = R_I \delta_P,&\theta'=0,\nonumber\\
&\xib = \xib_P = 2\frac{R_P^2}{a^2}-1=1+2\delta_P, &\etab= 1,\nonumber\\[0.2cm]
&\mathrm{where~~} \delta_P = \frac{R_P^2}{a^2}-1.
\end{align*}
Note that the adimensional parameter $\delta_P >0$ becomes small when the dipole is close to the surface.
The self-field $\bm{E}_{\mathrm{sf}-{\bot}}$ (at the dipole position) is given by
\begin{align}
\bm{E}_{\mathrm{sf}-{\bot}}&=E_0 \beta_\infty \left(  \frac{2}{\delta_P^3} + \frac{\epsr}{\epsr+1} \frac{1}{\delta_P^2}\left[ 
1 - \frac{1}{\delta_P(1+\delta_P)}\sum_{n=0}^\infty(2n+1)(n+1)c_n[\xib_P Q_n(\xib_P)-Q_{n+1}(\xib_P)] \right]\right)\hat{\bm{z}}
\end{align}
from which we deduce the modified decay rate as
\begin{align}
\frac{\Gamma_\bot}{\Gamma_0} = 1 + \frac{3}{2(k_1 a)^3}\mathrm{Im}\left\lbrace  \beta_\infty \left( \frac{2}{\delta_P^3} + \frac{\epsr}{\epsr+1} \frac{1}{\delta_P^2}\left[ 1 - \frac{1}{\delta_P(1+\delta_P)}\sum_{n=0}^\infty(2n+1)(n+1)c_n[\xib_P Q_n(\xib_P)-Q_{n+1}(\xib_P)] \right]\right)\right\rbrace
\label{QP decay rate}
\end{align}

It is worth noting that the coefficients $c_n$ in the series also depend on $\epsr$ and contribute to the material-dependence of the whole expression.

\section{Parallel dipole near a sphere}

In this section, we adapt the derivations presented for the perpendicular dipole
to the case of a parallel dipole. The main difference is that all spherical and spheroidal harmonic expansions
now contain Legendre functions $P_n^m,~Q_n^m$ with $m=1$, instead of $m=0$ for perpendicular dipoles.

\subsection{Solution in spherical coordinates}

The standard solution of this problem consists of expanding the dipole potential as a series of regular solid harmonics centred at the origin.
For a dipole along $x$, located on the $z$-axis at $z=R_P$, we have (analogous to Eq.~\ref{EqnPhiDip1}):
\begin{align}
\bar{\phi}_{\mathrm{dip}-{\parallel}} = \frac{a R_P x}{|\mathbf{r} - R_P \hat{\mathbf{z}}|^3}
=-\frac{a}{R_P}\sum_{n=1}^\infty~\left(\frac{r}{R_P}\right)^nP^1_n(\cos\theta)\cos\phi  \qquad (r<R_P).
\end{align}
The solution of the problem outside the sphere ($r>a$) is then given by $\bar{\phi}_\mathrm{out-||} = \bar{\phi}_{\mathrm{dip}-{\parallel}} + \bar{\phi}_{{\parallel}}$, with the reflected potential $\bar{\phi}_{{\parallel}}$ given by \cite{1941Stratton}:
\begin{align}
\bar{\phi}_{{\parallel}} = \sum_{n=1}^\infty ~ \beta_n \left(\frac{R_I}{r}\right)^{n+1}  ~ P_n^1(\cos\theta)\cos\phi, 
\label{EqnPhiSca2}
\end{align}
where $\beta_n$ are the adimensional sphere polarizabilities as defined in Eq. 3.

The reflected electric field solution outside the sphere is then:
\begin{align*}
\mathbf{E}_{\parallel}=E_0 \sum_{n=1}^\infty \beta_n \left(\frac{R_I}{r}\right)^{n+2}
\left[  (n+1)P_n^1(\cos\theta)\cos\phi\unitr 
- \frac{\dif}{\dif \theta}\left[  P_n^1(\cos\theta) \right]\cos\phi \unitt + \frac{P_n^1(\cos\theta)}{\sin\theta}\sin\phi\unitf \right]
\end{align*}
where $E_0 = p/(4\pi\epsilon_0 \epsilon_1 a^3)$. 
To calculate the self-field $E_{\mathrm{sf}-{\parallel}}$ (at $r=R_P$, $\theta=0$), we need to take the limit as $\theta\rightarrow 0$. We use the equalities:
\begin{align}
P_n^1(1)=0,\quad
\lim_{\theta\rightarrow 0}\left[\sin\theta{\frac{\dif}{\dif\theta} P_n^1(\cos\theta)}\right] = 
\lim_{\theta \rightarrow 0}\left[\frac{P_n^1(\cos\theta)}{\sin\theta}\right] = -\frac{n(n+1)}{2},
\end{align}
and
\begin{align}
\unitx = \sin\theta\cos\phi\unitr + \cos\theta\cos\phi\unitt-\sin\phi\unitf
\end{align}
to obtain 
\begin{align}
\mathbf{E}_{\mathrm{sf}-{\parallel}} &=E_0\sum_{n=1}^\infty \frac{n(n+1)}{2}\beta_n \left(\frac{a^2}{R_P^2}\right)^{n+2} \unitx.
\end{align}
The modified decay rate is therefore
\begin{align}
\frac{\Gamma_\parallel}{\Gamma_0}=1+\frac{3}{4(k_1 a)^3} \sum_{n=1}^\infty n(n+1)\mathrm{Im}(\beta_n) \left(\frac{a^2}{R_P^2}\right)^{n+2}.
\end{align}

\subsection{Analytic expressions for image terms}

We now come back to the potential (Eq. \ref{EqnPhiSca2}).
Following the same arguments as for the perpendicular dipole, one can recognize closed-form expressions for the first few dominant terms.
$\beta_n$ can be split and analytic expressions for the series can be identified.
Explicitly:
\begin{align}
\frac{n}{n(\epsilon+1)+1} = \frac{1}{\epsr+1} -\frac{1}{n(\epsr+1)^2}+\frac{1}{n(\epsr+1)^2(n\epsr+n+1)}.
\end{align}
The first term results in the series
\begin{align}
\sum_{n=1}^\infty ~ \left(\frac{R_I}{r}\right)^{n+1} P_n^1(\cos\theta)\cos\phi = - R_I^2\frac{x'}{r'^3}
\end{align}
where we have recognized the expansion of a dipole offset along the $z$-axis (and located at $z=R_I$).
This is the image dipole, whose orientation is in this case opposite to the real dipole.

The second term gives the series:
\begin{align}
\sum_{n=1}^\infty\frac{1}{n} \left(\frac{R_I}{r}\right)^{n+1}  P_n^1(\cos\theta)\cos\phi
\end{align}
In contrast with the case of a perpendicular dipole where an image point charge was identified,
it is not straightforward here to recognize an analytic expression.
To develop this further, we will use (for $n \ge 1$):
\begin{align}
P_n^1(\cos\theta) =  \frac{n}{\sin\theta} [\cos\theta P_n(\cos\theta)-P_{n-1}(\cos\theta)].
\end{align}
We then have
\begin{align}
\sum_{n=1}^\infty\frac{1}{n} \left(\frac{R_I}{r}\right)^{n+1}  P_n^1(\cos\theta)\cos\phi
&=\frac{\cos\phi}{\sin\theta}\sum_{n=1}^\infty\left(\frac{R_I}{r}\right)^{n+1}(\cos\theta P_n-P_{n-1}) \nonumber\\
&=\frac{\cos\phi}{\sin\theta}\left[\cos\theta\sum_{n=1}^\infty\left(\frac{R_I}{r}\right)^{n+1}P_n - \frac{R_I}{r}\sum_{n=0}^\infty\left(\frac{R_I}{r}\right)^{n+1}P_n \right] \nonumber\\
&=\frac{\cos\phi}{\sin\theta}\left[\cos\theta\left(\frac{R_I}{r'}-\frac{R_I}{r}\right) - \frac{R_I}{r}\frac{R_I}{r'} \right] \qquad \mathrm{using~Eq.~\ref{gf_formula}}\nonumber \\
&=\frac{R_I\cos\phi}{r\sin\theta}(\cos\theta'-\cos\theta)\qquad\mathrm{~~~~since~} \cos\theta' = \frac{r\cos\theta-R_I}{r'} \nonumber\\
&=\frac{R_I x}{\rho^2}(\cos\theta'-\cos\theta)
\end{align}
Note that the above expression is singular on the line from O to I, but converges to a finite value elsewhere even on the $z$ axis.

Putting those results together, and using $x'=x$, we have for the potential:
\begin{align}
\bar{\phi}_{\parallel}=-\beta_\infty \frac{R_I^2 x}{r'^3}
-\frac{\beta_\infty}{\epsr+1} \frac{R_I x (\cos\theta'-\cos\theta)}{r^2\sin^2\theta}
+\frac{\beta_\infty}{\epsr+1}\cos\phi\sum_{n=1}^\infty\left(\frac{R_I}{r}\right)^{n+1}\frac{P^1_n(\cos\theta)}{n(n(\epsr+1)+1)}
\end{align}

\subsection{New approach with spheroidal harmonics}

The main difference with the perpendicular dipole is that only the first dominant analytic term in the expression above is a point singularity. Since the second term exhibits the line singularity from O to I, there is no reason to isolate it when looking for the expansion in spheroidal harmonics.
We therefore start from the potential with the image dipole term only, namely:
\begin{align}
\bar{\phi}_{\parallel}=-\beta_\infty \frac{R_I^2 x}{r'^3}
-\beta_\infty\cos\phi \sum_{n=1}^\infty\left(\frac{R_I}{r}\right)^{n+1}\frac{P^1_n(\cos\theta)}{n(\epsr+1)+1}
\end{align}

This can then be converted into a spheroidal solid harmonics expansion using Eq.~\ref{PnmvsQPnm} for $m=1$, which reads:
\begin{align}
\left(\frac{R_I}{r}\right)^{n+1}P^1_n(\cos\theta) = -2\sum_{k=n}^\infty\frac{(-)^{n+k}}{n!(n-1)!}\frac{(k+n)!}{(k-n)!}\frac{2k+1}{k(k+1)}Q_k^1(\xib)P_k^1(\etab)
\end{align}
The series in the previous equation for $\bar{\phi}_{\parallel}$ then becomes
\begin{align}
\sum_{n=1}^\infty\frac{P^1_n(\cos\theta)}{n(\epsr+1)+1}\left(\frac{R_I}{r}\right)^{n+1}  &=
-2\sum_{n=1}^\infty\frac{1}{n\epsr+n+1}\sum_{k=n}^\infty\frac{(-)^{n+k}}{n!(n-1)!}\frac{(k+n)!}{(k-n)!}\frac{2k+1}{k(k+1)}Q_k^1(\xib)P_k^1(\etab) \nonumber\\
&=-2\sum_{n=1}^\infty\frac{2n+1}{n(n+1)}\sum_{k=1}^n\frac{(-)^{n+k}~k}{k\epsr+k+1}\frac{(n+k)!}{k!^2(n-k)!}Q_n^1(\xib)P_n^1(\etab)
\end{align}
In the last step, the order of summation was swapped, then the indices relabeled ($n\leftrightarrow k$). \\
Using the decomposition
\begin{align}
(\epsr+1)\frac{k}{k\epsr+k+1} + \frac{1}{k\epsr+k+1} = 1, 
\end{align}
we obtain the following relation:
\begin{align}
(\epsr+1)\sum_{k=0}^n\frac{(-)^{n+k}~k}{k\epsr+k+1}\frac{(n+k)!}{k!^2!(n-k)!} + \sum_{k=0}^n\frac{(-)^{n+k}}{k\epsr+k+1}\frac{(n+k)!}{k!^2!(n-k)!}= 
\sum_{k=0}^n (-)^{n+k}\frac{(n+k)!}{k!^2(n-k)!} = 1.
\end{align}
The latter equality can be obtained by evaluating Eq. \ref{PPnmvsPnm} at $m=0,~\xib=\etab=1$.
The second term above can be identified as $c_n$ (Eq. 14) and the sum in the
first term can start at $k=1$ without affecting the result, so we have:
\begin{align}
\sum_{k=1}^n\frac{(-)^{n+k}~k}{k\epsr+k+1}\frac{(n+k)!}{k!^2!(n-k)!}=\frac{1-c_n}{\epsr + 1}.
\end{align}
The potential therefore takes the form:

\begin{align}
\bar{\phi}_{\parallel}=-\beta_\infty \frac{R_I^2 x}{r'^3}
-\frac{2\beta_\infty}{\epsr+1}\cos\phi \sum_{n=1}^\infty \frac{2n+1}{n(n+1)} (c_n-1) Q_n^1(\xib)P_n^1(\etab).
\label{EqnPara}
\end{align}

\subsection{Electric field and modified decay rate with new approach}

Starting from the proposed new formula for the potential solution (Eq. \ref{EqnPara}), we can deduce the electric field in terms of the
more convergent spheroidal harmonics expansions:
\begin{align}
E_{\parallel}^r= & -E_0 \cos\phi\beta_\infty\left[ \frac{R_I^3 \sin\theta (2r^2-R_I z-R_I^2)}{r'^5}  \right.\nonumber\\
&\left. +  \frac{2}{\epsr+1}\frac{R_I}{r}\sum_{n=1}^\infty (2n+1)(c_n-1) \left(  \frac{Q_n^1(\xib)P_n^1(\etab)}{n} + \frac{Q_n^1(\xib)P_{n+1}^1(\etab)-Q_{n+1}^1(\xib)P_n^1(\etab)}{(n+1)(\xib-\etab)}  \right)  \right]
\nonumber\\[0.5cm]
E_{\parallel}^\theta =& -E_0 \cos\phi \beta_\infty\left[ \frac{R_I^3}{r'^5}\left(r'^2 \cos\theta -3 R_I r \sin^2\theta\right) \right.\nonumber\\
&\left. -\frac{2}{\epsr+1}\frac{R_I}{r\sin\theta} \sum_{n=1}^\infty (2n+1)(c_n-1)  \left(   \frac{Q_n^1(\xib)P_n^1(\etab)}{n}\cos\theta - \frac{R_I^2}{4rr'}\frac{Q_n^1(\xib)P_{n+1}^1(\etab)(\xib^2-1) - Q_{n+1}^1(\xib)P_n^1(\etab)(\etab^2-1) }{(n+1)}  \right)\right]
\nonumber\\[0.5cm]
E_{\parallel}^\phi = & -E_0\sin\phi\beta_\infty\left[\frac{R_I^3}{r'^3}+\frac{2}{\epsr+1}\frac{R_I}{r\sin\theta}\sum_{n=1}^\infty\frac{2n+1}{n(n+1)}(c_n-1)Q_n^1(\xib)P_n^1(\etab)    \right].
\end{align}

To calculate the self-field we take limits as $\theta\rightarrow0$ to obtain:
\begin{align}
\mathbf{E}_{\mathrm{sf}-{\parallel}}&=-E_0\beta_\infty \left\lbrace  \frac{1}{\delta_P^3} - \frac{2}{\epsilon+1}\sum_{n=1}^\infty(2n+1)(c_n-1)\frac{Q_n^1(\xib_P) }{\sqrt{\xib_P^2-1}}   \right\rbrace~ \hat{\bm{x}}
\end{align}
from which we deduce the modified decay rate as (using $\delta_P$ defined earlier)
\begin{align}
\frac{\Gamma_\parallel}{\Gamma_0} = 1 - \frac{3}{2(k_1 a)^3}\mathrm{Im}\left\lbrace \beta_\infty\left(\frac{1}{\delta_P^3} - \frac{2}{\epsr+1}\frac{1}{\sqrt{\delta_P(1+\delta_P)}}\sum_{n=1}^\infty(2n+1)(c_n-1){Q_n^1(\xib_P)}\right)   \right\rbrace.
\end{align}

The comparison of the convergence of this series to the spherical harmonic series is very similar to that obtained for the perpendicular dipole case.

\section{Definition and computation of the Legendre functions of the first and second kind}
\label{SecLegDef}

The associated Legendre functions of the first kind are widely used, but there are different conventions. We used the following definition (for $m \ge 0$):
\begin{align}
P_n^m(x)=
\begin{cases} 
      (-)^m (1-x^2)^{m/2} \partial_xP_n(x) & |x|<1 \\
      (x^2-1)^{m/2} \partial_xP_n(x) & |x|>1
\end{cases} 
\end{align}
They are computed by forward recurrence on $n$ (for a fixed $m$) using the relation:
\begin{align}
(n-m+1)P_{n+1}^m(x) = (2n+1)xP_n^m(x)-(n+m)P_{n-1}^m(x), \label{PnmRec}
\end{align}
and the initial conditions
\begin{align}
P_{m}^m (x) = \begin{cases}  (-)^m (2m-1)!!(1-x^2)^{m/2}& |x|<1\\
(2m-1)!!(x^2-1)^{m/2}& |x|>1
\end{cases} 
,\qquad P_{m+1}^m(x) = (2m+1)xP_{m}^m (x).
\end{align}
The associated Legendre functions of the second kind are much less common.
They are defined for $\xi \ge 1$ as follows \cite{2000Jansen}.
First, for $m=0$, we define:
\begin{align}
Q_n(\xi) = \frac{1}{2}\int_{-1}^{1} \frac{P_n(x)}{\xi-x} \dif x.
\end{align}
This gives for the first orders:
\begin{align}
Q_0(\xi) = \frac{1}{2}\ln\frac{\xi+1}{\xi-1},\qquad Q_1(\xi)=\xi Q_0(\xi) - 1,
\end{align}
We then define for $\xi \ge 1$ and $m\ge 0$, similarly to the functions of the first kind:
\begin{align}
Q_n^m(\xi)=(\xi^2-1)^{m/2}\frac{\dif^m Q_n}{\dif \xi^m}(\xi)
\end{align}

$Q_n^m$ obeys exactly the same recurrence as in Eq. \ref{PnmRec}. However,
stable computation of $Q_n(\xi)$ is more complicated as this simple forward recurrence becomes quickly numerically unstable as $n$ increases.
Instead, we have therefore used a stable backward recurrence described in Ref. \cite{Qalgorithm}, which is based on the modified Lentz algorithm.

While $Q_n^1$ can be calculated using the backward recurrence described above, alternatively it can be easily derived from the $m=0$ functions:
\begin{align}
Q_n^1(\xi) = \frac{n}{\sqrt{\xi^2-1}}\left[\xi Q_n(\xi) - Q_{n-1}(\xi)\right]. \qquad (n \ge 1)
\end{align} 
Note this also applies to $P_n^1$.
\bibliography{Tmatrix}

\end{document}